# WAVE MOTIONS IN MOLECULAR CLOUDS: RESULTS IN TWO DIMENSIONS

Curtis S. Gehman, Fred C. Adams
Marco Fatuzzo[†], and Richard Watkins

*Physics Department, University of Michigan*
*Ann Arbor, MI 48109*

[†]*current address: Wesleyan College*
*Macon, GA 31297*

## ABSTRACT

We study the linear evolution of small perturbations in self-gravitating fluid systems in two spatial dimensions; we consider both cylindrical and cartesian (i.e., slab) geometries. The treatment is general, but the application is to molecular clouds. We consider a class of equations of state which heuristically take into account the presence of turbulence; in particular, we consider equations of state which are *softer* than isothermal. We take the unperturbed cloud configuration to be in hydrostatic equilibrium. We find a class of wave solutions which propagate along a pressure supported cylinder (or slab) and have finite (trapped) spatial distributions in the direction perpendicular to the direction of propagation. Our results indicate that the dispersion relations for these two dimensional waves have similar forms for the two geometries considered here. Both cases possess a regime of instability and a fastest growing mode. We also find the (perpendicular) form of the perturbations for a wide range of wavelengths. Finally, we discuss the implications of our results for star formation and molecular clouds. The mass scales set by instabilities in both molecular cloud filaments and sheets are generally much larger than the masses of stars. However, these instabilities can determine the length scales for the initial conditions for protostellar collapse.

*Subject headings:* hydromagnetics – wave motions – interstellar: molecules – stars: formation

# 1. INTRODUCTION

It is widely believed that most current galactic star formation occurs in molecular clouds (cf. Shu, Adams, & Lizano 1987 for a review). Observations of these clouds reveal rich and complex structure (e.g., Myers 1991; Blitz 1993). In particular, filamentary and sheetlike structures are common, which in turn exhibit a large degree of substructure of their own (e.g., Schneider & Elmegreen 1979; de Geus, Bronfman, & Thaddeus 1990; Houlahan & Scalo 1992; Wiseman & Adams 1994). Clumps and cores that provide locations for new stars to form are often found along filaments and sheets. Thus, the study of the evolution of molecular clouds and their substructure is important for understanding the processes involved in present day star formation.

In this paper, we continue our study of the fluid dynamics of self-gravitating systems with an emphasis on wave motions in molecular clouds. In previous papers, we have shown that linear and nonlinear volume density waves can exist in molecular clouds and may be important in determining their substructure (Adams & Fatuzzo 1993; Adams, Fatuzzo, & Watkins 1993, 1994). This previous work has focused on the case of one spatial dimension (i.e., waves in a uniform medium). In this paper, we generalize our treatment to study wave motions in two spatial dimensions. In particular, we study waves propagating down the axes of molecular cloud filaments and sheets.

Most previous work has focused on the case of wave motions and instabilities in *uniform density* fluids (from Jeans 1928 to Dewar 1970, Langer 1978, Pudritz 1990). Some previous work on wave motions in isothermal molecular cloud filaments and sheets has been done; this work begins with clouds in hydrostatic equilibrium and hence nonuniform density. Many observations show clumps which appear nearly equally spaced along filaments (e.g., Schneider & Elmegreen 1979; McBreen et al. 1979; Dutrey et al. 1991); this finding has led to the idea that the clumps may arise from a gravitational instability with a particular length scale. Less frequently, it has been proposed that the clumps may be peaks of density waves which propagate along the filament. In the linear regime, both cases are treated by a linear perturbation analysis. Larson (1985) gives a good review of the early progress in doing this type of analysis. More recent work includes Nagasawa (1987), who performed such calculations for an idealized isothermal filament, including an axial magnetic field. Other workers (Nakamura, Hanawa, & Nakano 1991, 1993; Matsumoto, Nakamura, & Hanawa 1994) have added further embellishments to this model (e.g., rotation). Until now, however, these studies have typically neglected the effects of turbulence or non-isothermal equations of state. In this present work, we consider a *general barotropic equation of state* of the form $p = P(\rho)$. We concentrate this present discussion on equations of state which are softer than isothermal, since these may be the most relevant for molecular clouds (see §2).

A secondary motivation for this work is to generalize the current theory of star formation, which typically begins with spherically symmetric clouds (e.g., Larson 1972; Shu 1977; Terebey, Shu, & Cassen 1984). Models of star formation built upon these collapse calculations are reasonably successful and predict spectral energy distributions of forming stars that are in agreement with observed protostellar candidates (Adams, Lada, & Shu 1987; Butner et al. 1991; Kenyon, Calvet, & Hartmann 1993). However, departures from spherical symmetry are expected to occur on larger spatial scales, and the collapse of isothermal self-gravitating sheets (Hartmann et al. 1994) and filaments (Inutsuka & Miyama 1993; Nakamura, Hanawa, & Nakano 1995) has begun to be studied. The calculations of this paper help determine the length scales for producing initial perturbations in molecular cloud sheets and filaments.

This paper is organized as follows. In §2, we present the basic equations and discuss generalizations of the isothermal equation of state. In §3, the equilibrium solutions of interest are presented and discussed. In §4, we present a general linear perturbation analysis followed by specific results for the filament and the slab for various equations of state. We conclude, in §5, with a discussion and summary of our results.

## 2. GENERAL FORMULATION

In this section, we present the basic equations used to describe self-gravitating fluids, such as molecular clouds. We also discuss an equation of state of particular interest, i.e., a generalized (non-isothermal) barotropic equation of state that heuristically includes the effects of turbulence in molecular clouds.

In dimensionless units (see Appendix A), the equations of fluid dynamics with self-gravity can be written in the form

$$\frac{\partial \rho}{\partial t} + \nabla \cdot (\rho \mathbf{u}) = 0 \,, \tag{2.1}$$

$$\frac{\partial \mathbf{u}}{\partial t} + (\mathbf{u} \cdot \nabla)\mathbf{u} + \frac{1}{\rho}\nabla p + \nabla \psi = 0 \,, \tag{2.2}$$

$$\nabla^2 \psi = \rho \,, \tag{2.3}$$

where $\rho$ is the density, $\mathbf{u}$ is the velocity, and $\psi$ is the gravitational potential. The pressure $p$ is taken to have a barotropic form, i.e., the pressure is a function of the density only,

$$p = P(\rho) \,. \tag{2.4}$$

Most previous theoretical studies of molecular clouds have assumed an isothermal equation of state, $p = c_s^2 \rho$, where $c_s$ is the *isothermal sound speed*. We generalize this equation of state by adding a term which attempts to model the "turbulence" observed in molecular clouds (see, e.g., Lizano & Shu 1989). The equation of state takes the form

$$p = c_s^2 \rho + p_0 \log(\rho/\hat{\rho}) \,, \tag{2.5}$$

where $p_0$ is a constant, which may be determined empirically, and $\hat{\rho}$ is an arbitrary reference density (see Appendix A). In this paper, we shall always take our dimensionless variable $\rho$ to be 1 at the center of equilibrim configurations, so that $\hat{\rho} = \rho_c$, the central density. The logarithmic nature of the turbulence term arises from the empirical linewidth-density relation $\Delta v \propto \rho^{-1/2}$ (Larson 1981; Myers 1983; Dame et al. 1986;

Myers & Fuller 1992). After transforming to dimensionless variables, this equation of state can be written

$$p = P(\rho) = \rho + \kappa \log(\rho),  \qquad (2.6)$$

where we have defined

$$\kappa \equiv p_0/c_s^2 \hat{\rho}, \qquad (2.7)$$

which we refer to as the *turbulence parameter*. Typically, we are interested in clouds with densities of about 1000 cm$^{-3}$ and thermal sound speed $c_s \sim 0.20$ km/s. Observational considerations suggest that $p_0$ ranges from 10 to 70 picodyne/cm$^2$ (Myers & Goodman 1988; Solomon et al. 1987). Thus, molecular clouds are expected of have values of the turbulence parameter in the range $6 < \kappa < 50$. In this work, we want to include both the isothermal limit and the pure "logatropic" limit where the turbulence dominates, so we use the expanded range $0 \leq \kappa < \infty$.

Before leaving this section, we note that there is some theoretical motivation for considering equations of state which are softer than isothermal. It has been shown that when a molecular cloud begins to collapse (on large spatial scales), a wide spectrum of small scale wave motions can be excited (Arons & Max 1975; see also Elmegreen 1990). Additional energy input and wave excitation can also be produced by outflows from forming stars (Norman & Silk 1980). Because the clouds are supported by magnetic fields, these wave motions generally take the form of magnetoacoustic and Alfvén waves. In any case, these small scale wave motions have velocity perturbations which vary with the gas density in rough agreement with the observations described above. These wave motions can be modeled with an effective equation of state which is softer than isothermal (see Fatuzzo & Adams 1993; McKee & Zweibel 1995).

## 3. STATIC EQUILIBRIUM

In this section, we present and discuss the static equilibrium solutions that we will use in our subsequent linear perturbation analysis. We consider the unperturbed state of the system to be one of hydrostatic equilibrium ($\mathbf{u} \equiv 0$). Thus, the unperturbed state must satisfy the usual equation,

$$\frac{1}{\rho}\nabla^2 p - \frac{1}{\rho^2}\nabla p \cdot \nabla \rho + \rho = 0, \qquad (3.1)$$

where we have combined the force equation [2.2] and the Poisson equation [2.3]. In the following subsections, we find equilibrium solutions for both cylindrical and cartesian geometries.

### 3.1 Hydrostatic Equilibrium for The Filament

For the filament, we adopt cylindrical coordinates $(r, \phi, z)$ and assume azimuthal symmetry. Then the equilibrium equation becomes

$$\frac{d^2\rho_0}{dr^2} + \frac{1}{r}\frac{d\rho_0}{dr} + \left[\frac{P''(\rho_0)}{P'(\rho_0)} - \frac{1}{\rho_0}\right]\left(\frac{d\rho_0}{dr}\right)^2 + \frac{\rho_0^2}{P'(\rho_0)} = 0, \qquad (3.2)$$

where the primes denote derivatives with respect to density,

$$P'(\rho_0) \equiv [dp/d\rho]_{\rho=\rho_0},$$

and so on. In the isothermal case $p = P(\rho) = \rho$, the equilibrium solution is well known (Ostriker 1964) and has the simple form

$$\rho_0(r) = \left(1 + r^2/8\right)^{-2}. \tag{3.3}$$

For equations of state which include turbulence ($\kappa \neq 0$), we find the solutions numerically.

In Figure 1, we show the cylindrical static equilibrium solutions for various equations of state, i.e., for various values of the turbulence parameter $\kappa$. Figure 1 shows two interesting trends. The first is that the filament becomes wider as the turbulence contribution is increased. This behavior is expected and simply reflects the fact that greater pressure can support more mass. The second trend is that the *shape* of the density profile also changes as the turbulence parameter increases. In particular, the asymptotic behavior of the isothermal equilibrium and that of the turbulent equilibria ($\kappa > 0$) have different forms. For large radii $r$, the isothermal equilibrium profile behaves like $r^{-4}$ (see equation [3.3]), whereas the turbulent equilibria behave as $r^{-1}$. It follows that the mass per unit length of the isothermal filament is finite, but that of the turbulent filament is infinite. For the isothermal equilibrium, the mass per unit length $\mu$ of the filament is given by

$$\mu = \int_0^\infty 2\pi r \left(1 + \frac{1}{8}r^2\right)^{-2} dr = 8\pi. \tag{3.4}$$

The corresponding integral for any equation of state with turbulence ($\kappa \neq 0$) diverges. In fact, the mass per unit length of a gaseous filament in hydrostatic equilibrium diverges for *any* equation of state softer than isothermal (see Appendix B). In practice, however, molecular cloud filaments have an outer boundary (at a finite radius $R$) determined by either pressure equilibrium with the background interstellar medium or by tidal effects. We expect the radius $R$ to be large enough that the outer boundary condition has little effect on the modes calculated in this paper. In any case, although the mass per unit length of these filaments does not actually diverge, it does become much larger than that of the isothermal case by a factor of $\sim R \approx 50$. These results on mass scales have important implications for star formation, as we discuss in §5.

### 3.2 Hydrostatic Equilibrium for The Slab

For the slab, we use cartesian coordinates $(x, y, z)$ and assume translational symmetry in the $\hat{y}$- and $\hat{z}$-directions. The equilibrium equation now becomes

$$\frac{d^2\rho_0}{dx^2} + \left[\frac{P''(\rho_0)}{P'(\rho_0)} - \frac{1}{\rho_0}\right]\left(\frac{d\rho_0}{dx}\right)^2 + \frac{\rho_0^2}{P'(\rho_0)} = 0. \tag{3.5}$$

For the isothermal case ($p = \rho$), this equation can be integrated analytically to obtain the well known solution (Spitzer 1942; Shu 1992)

$$\rho_0 = \mathrm{sech}^2\left(x/\sqrt{2}\right). \tag{3.6}$$

Again, for equations of state including turbulence, we find the equilibrium density profiles numerically.

In Figure 2, we show the profile of slab equilibria for various equations of state. As for the cylindrical case considered above, the density profiles become wider and change their shape as the turbulence parameter $\kappa$ increases. In this case, the isothermal equilibrium density profile has an exponential fall off at large distances (see equation [3.6]) and, hence, has a finite surface density $\sigma = 2^{3/2}$. When turbulence is included, the fall off is much slower. For the purely logatropic limit, the density profile has its asymptotic form given by the following transcendental equation,

$$2\rho^2 \log[1/\rho] = \frac{1}{x^2} \,. \tag{3.7}$$

Although this density profile falls slightly faster than $1/x$, the mass per unit area still diverges.

As in the cylindrical case considered above, a qualitative difference exists between the isothermal equilibrium configuration and that of the purely logatropic (turbulence dominated) case. In the case of the filament, however, the mass per unit length *diverges* for all equations of the state which are softer than isothermal. In contrast, for the hydrostatically supported slab, the mass per unit area (the surface density) *is finite* for all equations of state except the softest case of the logatrope (see Appendix C).

The equilibrium configurations with the turbulence parameter $\kappa \approx 10$ are the most physically realistic. For the purely isothermal case, both the cylindrical equilibrium (where $\rho \sim r^{-4}$ at large radii) and the slab equilibrium (where the density falls off exponentially at large $x$) are too narrow compared with actual molecular cloud structures. The thicker equilibria resulting from including turbulence ($\kappa \neq 0$) more closely resemble true clouds. However, the mass per unit length of the filament and the surface density of slab both diverge for the $\kappa \neq 0$ cases. In practice, the cloud structures have outer boundaries where the internal cloud pressure drops to that of the ambient interstellar medium. Thus, the mass per unit length (and surface density) are *large, but finite* in practice. We discuss this issue further in §5.

## 4. PERTURBATIONS AND WAVE SOLUTIONS

In this section, we study perturbations about the hydrostatic equilibrium density distributions. After presenting the basic perturbation analysis (§4.1), we present results for clouds with both cylindrical (§4.2) and cartesian (§4.3) geometries.

### *4.1 General Perturbation Analysis*

We now consider perturbations about the basic equilibrium states, $\rho_0$, found in the previous section. We let the subscript '1' denote first order quantities and find the following first order equations of motion for this system:

$$\frac{\partial \rho_1}{\partial t} + \nabla \cdot (\rho_0 \mathbf{u}_1) = 0 \,, \tag{4.1}$$

$$\frac{\partial \mathbf{u}_1}{\partial t} + \frac{1}{\rho_0}\nabla p_1 - \frac{\rho_1}{\rho_0^2}\nabla p_0 + \nabla \psi_1 = 0 \,, \tag{4.2}$$

$$\nabla^2 \psi_1 = \rho_1 \,. \tag{4.3}$$

In order to eliminate the velocity $\mathbf{u}_1$, we multiply the first order force equation [4.2] by $\rho_0$ and take the divergence. We then use the time derivative of the continuity equation [4.1] to remove the velocity from the problem. We thus obtain the first order equation of motion

$$-\frac{\partial^2 \rho_1}{\partial t^2} + \nabla^2 p_1 - \frac{\rho_1}{\rho_0}\nabla^2 p_0 - \frac{1}{\rho_0}\nabla \rho_1 \cdot \nabla p_0 + \frac{\rho_1}{\rho_0^2}\nabla \rho_0 \cdot \nabla p_0 + \rho_0 \rho_1 + \nabla \rho_0 \cdot \nabla \psi_1 = 0 \,, \tag{4.4}$$

where we have used the Poisson equation in obtaining the penultimate term. After some rearrangement, this first order perturbation equation becomes

$$-\frac{\partial^2 \rho_1}{\partial t^2} + \nabla^2 p_1 - \frac{1}{\rho_0}\nabla \rho_1 \cdot \nabla p_0 + 2\rho_0 \rho_1 + \nabla \rho_0 \cdot \nabla \psi_1 = 0 \,. \tag{4.5}$$

We are interested in finding solutions to equation [4.5] which correspond to waves propagating down the center of a cylinder or along the center of a slab. We choose the $\hat{z}$-direction to be the direction of wave propagation for both cases. On the other hand, the unperturbed equilibrium state (as given by the solution to equation [3.1]) will be a function of only the perpendicular coordinate (denoted as $r$ for the cylindrical case and $x$ for the Cartesian case). Notice that the coefficients appearing in the differential equation [4.5] depend only on the perpendicular coordinate. The equations are linear (in $\rho_1$ and $\psi_1$) and, with some care, can be separated (see Appendix D), i.e., we take

$$\rho_1 \equiv f(\varpi)g(z)h(t) \tag{4.6}$$

$$\psi_1 \equiv \phi(\varpi)g(z)h(t) \,, \tag{4.7}$$

where we have used $\varpi$ to represent the perpendicular coordinate (either $x$ or $r$, depending on the geometry). We can separate the Laplacian operator by defining

$$\nabla^2 \equiv \frac{\partial^2}{\partial z^2} + \nabla_\perp^2 \tag{4.8}$$

where we have defined a perpendicular Laplacian operator $\nabla_\perp^2$. For the case of cartesian coordinates, this operator is simply

$$\nabla_\perp^2 = \frac{\partial^2}{\partial x^2} \,, \tag{4.9a}$$

whereas for cylindrical coordinates it takes the form

$$\nabla_\perp^2 = \frac{\partial^2}{\partial r^2} + \frac{1}{r}\frac{\partial}{\partial r} \,. \tag{4.9b}$$

As shown in Appendix D, we obtain traveling wave solutions in the $\hat{z}$-direction, i.e.,

$$h(t) = h_0 e^{\pm i\omega t}, \tag{4.10}$$

$$g(z) = g_0 e^{\pm ikz}, \tag{4.11}$$

where $h_0$ and $g_0$ are constants. The equations for the dynamics in the remaining perpendicular direction take the form

$$\nabla_\perp^2 f + \left[2\frac{P''(\rho_0)}{P'(\rho_0)} - \frac{1}{\rho_0}\right]\frac{d\rho_0}{d\varpi}\frac{df}{d\varpi} + [P'(\rho_0)]^{-1}\frac{d\rho_0}{d\varpi}\frac{d\phi}{d\varpi} + V(\varpi)f = k^2 f, \tag{4.12}$$

$$\nabla_\perp^2 \phi - f = k^2 \phi, \tag{4.13}$$

where we have defined a function $V$:

$$V(\varpi) \equiv [P'(\rho_0)]^{-1}\left\{\omega^2 + 2\rho_0 + P'''(\rho_0)\left(\frac{d\rho_0}{d\varpi}\right)^2 + P''(\rho_0)\nabla_\perp^2\rho_0\right\}. \tag{4.14}$$

The problem is made complete by specifying the boundary conditions. We take

$$f = 1,\ \frac{df}{d\varpi} = 0,\ \frac{d\phi}{d\varpi} = 0,\ \text{at } \varpi = 0; \tag{4.15}$$

$$f = 0,\ \frac{d\phi}{d\varpi} = 0,\ \text{at } \varpi = \infty. \tag{4.16}$$

As formulated above, these equations constitute an eigenvalue problem with $k^2$ as the eigenvalue. There is an additional parameter in the problem, namely $\omega^2$. In practice, however, we choose the value of the wave number $k$ and find the value of the parameter $\omega^2$ that makes $k^2$ an eigenvalue.

As is usual in eigenvalue problems, there can be multiple eigenvalues $k^2$ corresponding to eigenfunctions with different numbers of nodes. Although $\omega^2$ is not strictly an eigenvalue, it can similarly have multiple values corresponding to eigenmodes for fixed $k^2$. When the points $(k, \omega^2)$ that correspond to eigenmodes are plotted, they naturally fall onto a set of curves that are the branches of the dispersion relation. We shall always be interested in the branch that has the smallest $\omega^2$ and corresponds to eigenfunctions with the fewest nodes. Note that it is not necessary for all modes on a particular branch to have the same number of nodes. As we shall see below, the dispersion relation we calculate has a minimum of $\omega^2$ occuring at non-zero $k$, so that near this minimum there are $\omega^2$ values that correspond to two values of $k$; modes to the right of the minimum have no nodes, while modes to the left have one node. One interesting result of this behavior is that one can determine whether the $k = 0$, $\omega^2 = 0$ mode is on the lowest branch by counting the number of nodes in its eigenfunction ($k = 0$, $\omega^2 = 0$ must be an eigenvalue in any case where there are a continuous set of equilibrium solutions with different central densities). If this mode has more than one node, then it is not on the lowest branch of the dispersion relation and there must exist at least one other mode with $k = 0$ and $\omega^2 < 0$.

Among the cases we consider, this situation obtains only for a filament with a turbulent equation of state (see Figure 3).

Before obtaining numerical solutions for this problem, we want to gain some basic intuition. We thus conceptually consider the simplest case in which the second and third terms in equation [4.12] are small, i.e., the equation of motion reduces to the form

$$\nabla^2_\perp f + V(\varpi)f = k^2 f \,. \tag{4.17}$$

This equation is, of course, simply the time independent Schrödinger wave equation (SWE). Although the quantity $-V$ plays the role of the potential, the analogy is not exact because an additional parameter $\omega^2$ appears in the definition [4.14] of $V$. We are interested in solutions for which the wave function $f$ decays to zero at large distances; these solutions correspond to bound states in the problem. We stress that this SWE analogy cannot be taken too far and that one must solve the full equation (as we do below); however, this analogy does provide a means of visualizing the dynamics of the problem.

Another way to conceptualize the dynamics is to consider the molecular cloud filament (or slab) to be somewhat like a wave guide. Modes with positive $k^2$ and $\omega^2$ correspond to waves which propogate in the $\pm\hat{z}$-direction but are otherwise trapped in the core of the filament or slab; in other words, the waves do not propagate in the perpendicular direction. This situation is analogous to light traveling in a fiber optic cable; trapping of the waves occurs due to total internal reflection, which in our case is a result of the change in the index of refraction that accompanies the change in density. Modes with negative $\omega^2$ correspond to linearly unstable perturbations. Again, this analogy is not exact, but it does provide a means of interpreting the problem.

### 4.2 Wave Propagation and Instabilities along a Filament

In this section, we consider waves propagating along filamentary molecular cloud structures. In particular, we adopt a cylindrical coordinate system $(r, \phi, z)$ and assume axial symmetry. The waves are propagating along the axis of the cylinder in the $\hat{z}$-direction. We take $\varpi = r$ and the perpendicular Laplacian $\nabla^2_\perp$ is given by equation [4.9b].

After adopting a particular equation of state and a particular value of the wave number $k$, both the eigenfunctions, $f_k(r)$ and $\phi_k(r)$, of equations [4.12] and [4.13] and the corresponding parameter $\omega^2_k$ are determined numerically using a relaxation method (see Appendix E). By varying the values of the wave number $k$, we can construct a dispersion relation for each equation of state. Here, we focus on equations of state of the form given by equation [2.6]; we thus have a one parameter family of equations of state (where we can vary the turbulence parameter $\kappa$). The limit of vanishing turbulence parameter $\kappa \to 0$ corresponds to a purely isothermal equation of state. The opposite limit, $\kappa \to \infty$, corresponds to a purely logatropic equation of state. The intermediate range $\kappa \sim 6 - -50$ is the most physically relevant.

In Figure 3, we show the dispersion relations for various equations of state. We have scaled the wave number to account for the difference between the isothermal sound speed, $c_s$, which was used in the transformation to dimensionless variables, and the *effective* sound speed at the center of the filament,

$$c_{\text{eff}} = [dp/d\rho]_{r=0}^{1/2} = (1+\kappa)^{1/2}.$$

Notice that for values of $\kappa$ larger than about 3, the dispersion relation quickly converges to the limiting case of pure turbulence, which corresponds to $\kappa \to \infty$. We also note that our dispersion relation for the pure isothermal case agrees well with that obtained previously (Nagasawa 1987) in a similar calculation.

For all equations of state considered here, there exists a region of instability, $0 < k < k_{\text{crit}}$, where $\omega_k^2 < 0$. For these unstable modes, the magnitude of $|\omega_k|$ determines the growth rate. Since each dispersion relation exhibits a minimum value of $\omega^2$, say at $k = k_{\text{fast}}$, there exists a *fastest growing mode* for each given equation of state. The wavelength of this fastest growing mode is simply $\lambda_{\text{fast}} = 2\pi/k_{\text{fast}}$. This wavelength represents the length scale of the perturbations which grow the fastest; thus, we anticipate that clumps forming along a filament due to gravitational instability will have a length scale close to $\lambda_{\text{fast}}$. For example, for the isothermal filament we find

$$\lambda_{\text{fast}}^{\text{isotherm}} \approx 0.777 \, \text{pc} \left[\frac{c_s}{0.20 \, \text{km/s}}\right] \left[\frac{\rho_c}{4 \times 10^{-20} \, \text{g/cm}^3}\right]^{-1/2}, \quad (4.18\text{a})$$

where $c_s$ is the thermal sound speed, and $\rho_c$ is the equilibrium central density. For the purely logotropic case,

$$\lambda_{\text{fast}}^{\text{log}} \approx 1.95 \, \text{pc} \left[\frac{\hat{p}}{4 \times 10^{-11} \, \text{dyne/cm}^2}\right]^{1/2} \left[\frac{\rho_c}{4 \times 10^{-20} \, \text{g/cm}^3}\right]^{-1}. \quad (4.18\text{b})$$

Unlike in the isothermal case, for the turbulent filament there is a timescale associated with collapse of the filament as a whole. This timescale can be calculated using the growth rate for the $k = 0$ mode (see Figure 3). Since this timescale is generally larger than that of the fastest growing mode, we shall not be concerned with this issue here.

In the isothermal case, this length scale of fragmentation leads to a mass scale for the fragmentation, since the mass per unit length of the equilibrium state is finite. We find the fragmentation mass scale to be

$$M_{\text{frag}} = \mu \lambda_{\text{fast}}^{\text{isotherm}} \approx 14.5 \, \text{M}_\odot \left[\frac{c_s}{0.20 \, \text{km/s}}\right]^3 \left[\frac{\rho_c}{4 \times 10^{-20} \, \text{g/cm}^3}\right]^{-1/2}, \quad (4.19)$$

where $\rho_c$ is the central density of the isothermal equilibrium. Notice that this mass scale is still much larger than that of a typical star ($M_* \sim 0.5 \, M_\odot$). The central density we have used here corresponds to a number density $n \sim 10^4 \, \text{cm}^{-3}$, a typical value for the high density regions of molecular clouds (see, e.g., the reviews of Blitz 1993; Myers 1991). Notice also that we expect the thermal sound speed to be greater than (or equal to) 0.20 km/s and hence this estimate represents a lower bound. Next we note that a corresponding mass scale cannot be constructed for the cases which include turbulence because the mass per unit length of the equilibrium state is formally infinite. Thus, the mass scales set by the fragmentation of molecular cloud filaments are generally much larger than the masses of stars.

A sample of the density eigenfunctions $f_k(r)$ is shown in Figure 4. They behave qualitatively as expected. The profiles are localized radially with widths roughly comparable to those of the equilibrium profiles, $\rho_0$. Notice that the shape of the eigenfunctions is nearly independent of the wave number $k$. In particular, both unstable modes (those with $\omega^2 < 0$) and propagating modes (those with $\omega^2 > 0$) have nearly the same shape. However, eigenfunctions with $k < k_{\text{fast}}$ have one node while those with $k > k_{\text{fast}}$ are positive definite. The eigenfunctions are only weakly dependent upon the equation of state (compare Figures 4a and 4b).

Cross-sectional images of perturbed filaments are shown in Figure 5. The images consist of isodensity contours and an array of arrows which indicate the velocity field in just half of an image. The fastest growing mode for the isothermal filament is depicted in Figure 5a; a propagating isothermal wave, in Figure 5b. Figures 5c and 5d show the corresponding perturbations for a filament with turbulence ($\kappa = 10$). Notice the distinct phase relationships between the density and velocity field for the propagating and the unstable modes (compare Figures 5a and 5b or Figures 5c and 5d). Also, notice the difference in shape of the clumps between the isothermal and the turbulent filaments (compare Figures 5a and 5c or Figures 5b and 5d).

Measuring the actual physical spacing, $\lambda_{\text{fast}}$, of clumps along a molecular cloud filament is rather difficult, since the uncertainty in the distance and orientation of the cloud is often large. A more easily measured quantity is the ratio

$$\mathcal{F} = \frac{\lambda_{\text{fast}}}{R_{\text{HWHM}}}, \qquad (4.20)$$

where $R_{\text{HWHM}}$ is the half-width at half-maximum of the filament. The dependence of this ratio on the turbulence parameter, $\kappa$, is shown in Figure 6. Notice that the ratio increases as the turbulence parameter is increased. The ratio $\mathcal{F}$ changes about 29% from the isothermal case to the limiting turbulent case, where $p = \log \rho$. This ratio also provides a measure of the elongation of the clumps (compare Figures 5a and 5c), though such a physical feature of the clumps may evolve significantly once the perturbation becomes non-linear.

In addition to increasing the length scale, turbulence also leads to an increase in the growth rate of the fastest growing mode. This finding is indicated by the the fact that, as the turbulence parameter $\kappa$ increases, the minimum value of $\omega^2(k)$ in the dispersion relation decreases. The dependence of the rate of growth $|\omega_{\text{fast}}|$ on the turbulence parameter $\kappa$ is shown in Figure 7. The growth rate for the limiting turbulent (logotropic) case is about 23% larger than for the isothermal case, where it is given by

$$|\omega_{\text{fast}}^{\text{isotherm}}| \approx 6.26 \times 10^{-14} \text{sec}^{-1} \left[\frac{\rho_c}{4 \times 10^{-20} \text{g/cm}^3}\right]^{1/2}. \qquad (4.21)$$

The corresponding time scale is given by $t \sim 1/\omega \sim 5 \times 10^5$ yr.

The dispersion relation obtained here can be compared to results from other theoretical models. One alternate way to model wave propagation along a filament is to consider one-dimensional waves in a uniform medium with a modified gravitational force that falls off with distance. (Normally, one-dimensional perturbations in a uniform medium produce gravitational forces that are independent of distance, while axial perturbations on a filament interact gravitationaly with forces that fall of roughly as the inverse square of the distance for large separations.) Yukawa potentials provide one means of realizing this behavior (Adams et al. 1994). In particular, we write the Poisson equation in the generalized form

$$\frac{\partial^2 \psi}{\partial x^2} = m^2 \psi + \rho \,. \tag{4.22}$$

The Green's function for the operator $\partial^2/\partial x^2 - m^2$ has an exponential fall off and hence produces an exponential fall off in the gravitational force between (one-dimensional) point masses. The value of the parameter $m$ determines the effective range of the force. The advantage of using such an approximation is to simulate the higher dimensional behavior of gravity while retaining a one-dimensional problem. As a result of this simplification, many nonlinear solutions and results can be found analytically for this model (see Adams et al. 1993, 1994).

The dispersion relation for linear waves can be easily obtained in this theory. In dimensionless units, we obtain the form

$$\omega^2 = \frac{\partial p}{\partial \rho} k^2 - \frac{k^2}{k^2 + m^2} \rho_0 \,. \tag{4.23}$$

In Figure 8, we show an analytic fit of this functional form to the numerical dispersion relation for the isothermal cylinder. We note that the Yukawa dispersion relation has the correct general form. However, the curvature of the function is *not* correct in the limit of small wave number $k \ll 1$. This result is expected because the Yukawa theory produces an exponential fall-off to the gravitational force at large distances (small wave numbers) and hence does not provide a good quantitative approximation. The Yukawa theory does provide a good approximation to the true two-dimensional theory at intermediate and large wave numbers, $k \gtrsim k_{\rm fast}$.

*4.3 Wave Propagation and Instabilities in the Slab*

In this section, we consider waves propagating along molecular cloud slabs. Adopting a cartesian coordinate system $(x, y, z)$, we assume that the cloud fluid is in hydrostatic equilibrium in the $\hat{x}$-direction and consider wave propagation in the $\hat{z}$-direction. The third ($y$) dimension is mathematically suppressed, and we are thus implicitly assuming that the slab has an infinite extent in the $\hat{y}$-direction. Here $\varpi = x$, and hence the perpendicular Laplacian $\nabla_\perp^2$ is given by equation [4.9a]. The problem is solved using the same numerical technique as used in the cylindrical case (again, see Appendix E).

The resulting dispersion relations for waves propagating through molecular cloud slabs are shown in Figure 9, where we have again scaled the wave number by the effective sound speed $c_{\text{eff}}$. Notice that these dispersion relations are similar in form to those of the filament discussed previously (compare Figures 3 and 9; for the isothermal case, see also Ledoux 1951, Simon 1965, and Larson 1985). As in the case of filaments, the dispersion relations for slabs exhibit a region of instability. Also, the rate of growth $|\omega|$ of the instability achieves a maximum at a finite length scale, and decreases as the wave number $k \to 0$ (i.e., as the wavelength $\lambda \to \infty$). Unlike the turbulent filament, however, the turbulent slab is marginally stable to perturbations of infinite wavelength. Thus, we need not be concerned that the turbulent slab-like cloud will collapse as a whole before finite instabilities grow significantly. We also note that the dispersion relations we have calculated are very similar (up to a scale factor) to that of an infinitesimally thin slab (see, e.g., Larson 1985); for a turbulent thin slab, the dispersion relation is given by $\omega^2 = (1 + \kappa)k^2 - \sigma k/2$.

For perturbations of the slab, we find that the wavelengths of the fastest growing modes for the isothermal and purely logotropic cases are given by

$$\lambda_{\text{fast}}^{\text{isotherm}} \approx 0.672\,\text{pc} \left[\frac{c_s}{0.20\,\text{km/s}}\right] \left[\frac{\rho_c}{4 \times 10^{-20}\,\text{g/cm}^3}\right]^{-1/2}, \qquad (4.24\text{a})$$

$$\lambda_{\text{fast}}^{\text{log}} \approx 1.32\,\text{pc} \left[\frac{\hat{p}}{4 \times 10^{-11}\,\text{dyne/cm}^2}\right]^{1/2} \left[\frac{\rho_c}{4 \times 10^{-20}\,\text{g/cm}^3}\right]^{-1}, \qquad (4.24\text{b})$$

For any equation of state stiffer than the logatropic case, a finite fragmentation mass scale exists for molecular cloud slabs or sheets. This mass scale is simply given by

$$M_{\text{frag}} = \pi \left(\frac{\lambda_{\text{fast}}}{2}\right)^2 \sigma = \frac{\pi^3 \sigma}{k_{\text{fast}}^2}, \qquad (4.25)$$

where $\sigma$ is the surface density of the slab and where $\lambda_{\text{fast}}$ is the wavelength of the fastest growing perturbation, and $k_{\text{fast}}$ is the corresponding wave number. For the isothermal slab we find

$$M_{\text{frag}} \approx 21.1\,M_\odot \left[\frac{c_s}{0.20\,\text{km/s}}\right]^3 \left[\frac{\rho_c}{4 \times 10^{-20}\,\text{g/cm}^3}\right]^{-1/2}. \qquad (4.26)$$

Notice that, once again, this mass scale is much larger than that of a typical star—by a factor of 40 for a star of 0.5 $M_\odot$. For equations of state softer than isothermal, this mass scale becomes even larger. In the logatropic limit, the surface density of the equilibrium configuration diverges and hence the mass scale of fragmentation becomes formally infinite.

## 5. DISCUSSION

In this paper, we have studied wave motions and instabilities in two spatial dimensions for self-gravitating astrophysical fluids. Although many of the results are general, the application to molecular clouds is our primary motivation. We have accounted for the presence of turbulence in molecular clouds by adding a logarithmic term to the equation of state. We have found the hydrostatic equilibrium states and studied wavelike perturbations. Specifically, we have numerically determined the dispersion relations and the structure of the perturbations.

### 5.1 Summary of Results

We have found several results which describe wave motions in molecular clouds and which add to our general understanding of fluid dynamics in self-gravitating physical systems. These results can be summarized as follows:

[1] We have studied the equilibrium configurations for self-gravitating filaments with various equations of state. The equilibrium configuration of the isothermal filament has a finite mass per unit length, whereas the mass per unit length *diverges* for any equation of state softer than isothermal (see Appendix B).

[2] We have also studied the equilibrium configurations of molecular cloud sheets with varying equations of state. In this case, the surface density (mass per unit area) is finite for all equations of state stiffer than that of the logatropic limit (see Appendix C). The surface density for the logatropic case diverges very slowly (slower than logarithmically).

[3] We have found the dispersion relations for waves in self-gravitating filaments and slabs. For perturbations of both the filament and the slab, the dispersion relations have the same general form for all equations of state considered here. The dispersion relations show that the square of the frequency $\omega^2$ is always positive at large wave numbers $k$, negative for small wave numbers, and obtains a minimum value for some intermediate wave number; this minimum implies the existence of a fastest growing wavelength. As the level of turbulence increases (i.e., as the equation of state becomes softer), both the length scale and the growth rate of the fastest growing instability increase.

[4] The ratio of the length scale of the fastest growing instability to the size (width) of the filament also increases as the turbulence parameter increases. This ratio grows by roughly 30% from the purely isothermal case to the purely turbulent limit ($\kappa \to \infty$). In the linear regime, this effect also implies that the shape or profile of the clumps will be elongated by the presence of turbulence.

[5] We have demonstrated that unstable and propagating perturbations show a qualitative difference in their phase relationships between the density field and the velocity field. In other words, the velocity fields of the perturbations show a different form for the propagating and unstable cases (compare Figures 5a and 5b with 5c and 5d, respectively). Thus, the velocity field can be used to determine whether molecular cloud structure is best described by wave motions or instabilities (see §5.2).

[6] We have compared the dispersions relations obtained from these two-dimensional calculations with those obtained from one dimensional "charge density" theories (Adams et al. 1994). These latter theories were developed to provide model equations which mimic the effects of higher dimensions while retaining the mathematical simplicity of a one dimensional problem. The dispersion relations for the filament are in reasonable agreement with those obtained from the charge density theory using a Yukawa potential;

this agreement supports the possibility that charge density theory will provide useful results for nonlinear waves in molecular cloud filaments.

[7] We have found the mass scales associated with the fastest growing instabilities in molecular cloud filaments and slabs. For filaments, the mass scale of fragmentation is given by equation [4.19] for an isothemal equation of state; the corresponding mass scale diverges for any softer equation of state. For the slab geometry, the mass scale of fragmentation is given by equation [4.26] for the isothermal case. This mass scale becomes larger as the equation of state becomes softer and diverges for a logatropic equation of state. In general, the mass scales set by the fastest growing modes in both filaments and slabs are much larger than the mass of a typical star.

The above results have important implications for the theory of star formation. In the current theory (cf. Shu et al. 1987), stars determine, in part, their own masses through the action of powerful winds and outflows. This theory thus assumes that the process of star formation has an unlimited (infinite) supply of material. In practice, the true supply of material will not be infinite, but will be much larger than the masses of the forming stars. However, star formation in isothermal molecular cloud filaments does have a finite mass scale, defined by the wavelength of the fastest growing instability and the mass per unit length (see equation [4.19]). Molecular clouds filaments which include turbulence have an infinite mass per unit length and hence an infinite supply of mass. Thus, the process of star formation can, in principle, be qualitatively different in isothermal filaments and those with softer equations of state. However, even in the isothermal case, the mass scale set by fragmentation is much larger than the mass of a typical star.

For the case of molecular cloud sheets, the surface density $\sigma$ is essentially finite for all cases of interest. Since the analysis of this paper shows that such sheets will be unstable for perturbations with finite length scale $\lambda$, a finite mass scale $M \sim \lambda^2 \sigma$ for perturbations (star formation) always exists (see, e.g., equation [4.26]). However, this mass scale is generally much larger than the mass scale of forming stars. Thus, the idea that stars, in part, determine their own masses remains applicable for star formation in molecular cloud sheets.

Although the instabilities studied in this paper do not directly determine the mass scales for star formation in molecular cloud filaments and sheets, they do determine (in part) the initial conditions for protostellar collapse. Thus, the results of this paper can be used as a starting point for collapse calculations in both cylindrical (Inutsuka & Miyama 1993; Nakamura et al. 1995) and sheetlike geometries (Hartmann et al. 1994).

*5.2 Comparison with Observed Molecular Clouds*

The waves and instabilities studied here can be compared with observations. In particular, the theory makes definite predictions that can be used to help understand molecular cloud structure.

The velocity field of the perturbations can be used to determine whether molecular cloud filaments (or slabs) contain propagating wave motions or instabilities. We have shown that the structure of the velocity field for propagating waves is strikingly different than that of growing instabilities (Figure 5). For unstable (growing) perturbations, the velocity vectors point inward toward the peak of the density field. For stable (propagating) modes, the velocity vectors do not converge at the peak of the density distribution. Although true observed velocity fields will look somewhat different than these figures due to projection effects, the basic difference between propagating and unstable modes should remain.

Observations should be able to determine the degree to which molecular cloud filaments depart from the purely isothermal model. In other words, we should be able to to determine which equation of state best describes molecular cloud filaments. The initial equilibrium configuration is quite different for the isothermal and non-isothermal equations of state used here. At large radial distances, the density distributions have the form $\rho \sim r^{-4}$ for the isothermal case and $\rho \sim r^{-2/(2-\Gamma)}$ for softer equations of state with the form $P \sim \rho^{\Gamma}$. For the purely logatropic limit, $\rho \sim r^{-1}$. Thus, the difference between an isothermal model and a logatropic model (the difference between $\rho \sim r^{-4}$ and $\rho \sim r^{-1}$) should be easily determined from observations.

Furthermore, the structure and appearance of the perturbations depend on the overall level of turbulence. As the level of turbulence in the filament increases (i.e., as the turbulence parameter $\kappa$ increases), the ratio of the clump spacing to the filament width (Figure 6) increases. The growth rate of the fastest growing instability (Figure 7) increases as well. Finally, turbulence increases the elongation of the clumps (Figure 5). The clump spacing, the filament width, and the degree of elongation of the clumps should all be observable with existing telescopes.

The most quantitative analysis of periodic structure in filaments has been provided by Schneider & Elmegreen (1979) and Dutrey et al. (1991). More is needed. The models developed here and other similar efforts produce results which will allow such observational work to determine the significance of turbulence and other physical processes in molecular clouds.

### 5.3 Directions for Future Work

Although we have begun to study wave motions in two dimensions in molecular clouds, many directions for future work remain. This future work includes both theoretical and observational studies. The observational work should test the theoretical predictions outlined in the previous subsection. In particular, measurements of the velocity fields in molecular cloud filaments will be particularly useful.

The turbulence observed in molecular clouds is often assumed to arise from small scale magnetohydrodynamic motions, such as Alfvén waves, within the clouds. In this work, we included the effects of such turbulence by incorporating a term in the equation of state which represents the effffective pressure due to turbulence. It is widely believed, however, that the large scale, time-averaged magnetic field also provides a significant fraction of the pressure support for molecular clouds. In fact, other studies (e.g., Nakamura et al. 1991, 1993) have studied waves and instabilities in the presence of magnetic fields, but have not included the effects of turbulence. In the future, both aspects of the problem (large scale magnetic fields and small scale turbulence) should be considered simultaneously.

Also, in future work, it will be interesting to investigate the behavior of perturbations assuming a wider variety of equations of state. For example, while we have focused here on equations of state that are softer than isothermal, stiffer equations of state should also be examined. For these equations of state, the equilibrium configurations of filaments have a finite radius, outside of which the density is zero. This type of structure can lead to qualitatively different types of perturbations; in particular, preliminary work has shown that it is possible to have modes which are trapped in the outer part of a filament (along the outer radius) instead of in its core.

Other functional forms for the equation of state of a turbulent gas should be considered as well. The logarithmic term used in this work was deduced empirically. Theoretical studies of turbulence in molecular clouds may lead to different formulations for the turbulent contribution to the pressure.

We note that the clumps observed in molecular cloud filaments are *not* small perturbations on an equilibrium filament. In general, molecular cloud substructure lies in the fully nonlinear regime. In our previous work, we began to study nonlinear waves in molecular clouds (e.g., Adams et al. 1993, 1994; see also Infeld & Rowlands 1990); however, this work was restricted to one spatial dimension. Ultimately, we hope to include nonlinear effects in our two dimensional study of structure in molecular clouds.

Acknowledgements

We would like to thank Phil Myers, Nathan Schwadron, Michael Weinstein, and Jennifer Wiseman for useful discussions. This work was supported by an NSF Young Investigator Award, NASA Grant No. NAG 5-2869, and by funds from the Physics Department at the University of Michigan; we also thank the Institute for Theoretical Physics at U. C. Santa Barbara and NSF Grant No. PHY94-07194.

## APPENDIX A: TRANSFORMATION TO DIMENSIONLESS VARIABLES

The Euler equations for the physical fields of a fluid can be written as

$$\frac{\partial \rho}{\partial t} + \nabla \cdot (\rho \mathbf{u}) = 0, \tag{A1}$$

$$\frac{\partial \mathbf{u}}{\partial t} + (\mathbf{u} \cdot \nabla)\mathbf{u} + \frac{1}{\rho}\nabla p + \nabla \psi = 0, \tag{A2}$$

$$\nabla^2 \psi = 4\pi G \rho, \tag{A3}$$

where $\rho$ is the density, $\mathbf{u}$ is the velocity, and $p$ is the pressure of the fluid and $\psi$ is the gravitational potential. In order to simplify the problem, we wish to transform to dimensionless variables. We perform the following transformation

$$\begin{aligned}
\mathbf{u} &\longrightarrow \hat{u}\mathbf{u} \\
\rho &\longrightarrow \hat{\rho}\rho \\
\psi &\longrightarrow \hat{u}^2 \psi \\
p &\longrightarrow \hat{u}^2 \hat{\rho} p \\
t &\longrightarrow \hat{x} t / \hat{u} \\
\mathbf{x} &\longrightarrow \hat{x}\mathbf{x}
\end{aligned} \tag{A4}$$

where $\hat{x} \equiv \hat{u}/(4\pi G \hat{\rho})^{1/2}$. Throughout this paper, we let $\hat{\rho} = \rho_c$, the central density of the equilibrium. For cases where the equation of state takes the form of equation [2.5] (including $\hat{p} = 0$), we let $\hat{u} = c_s$, the thermal sound speed. When we consider the equation of state of the form $p = \hat{p} \log(\rho/\hat{\rho})$, we set $\hat{u} = (\hat{p}/\hat{\rho})^{1/2}$. With the transformation [A4], the fluid equations [A1–A3] are cast into the form [4.1–4.3].

## APPENDIX B: MASS PER UNIT LENGTH OF FILAMENTS

In this Appendix, we show that the mass per unit length diverges for any equation of state which is softer than isothermal. We begin by considering a class of equations of state of the form

$$p = \mathrm{K}\rho^\Gamma . \tag{B1}$$

With this form, equations of state which are softer than isothermal correspond to values of $\Gamma < 1$. The equation for hydrostatic equilibrium of a filament can be written in the form

$$\frac{\partial}{\partial r}\left[r\rho^{\Gamma-2}\frac{\partial \rho}{\partial r}\right] + \alpha \rho r = 0 , \tag{B2}$$

where we have defined the constant $\alpha = 1/\mathrm{K}\Gamma$. Next, we define the effective power-law index $q$ of the density distribution,

$$q \equiv -\frac{r}{\rho}\frac{\partial \rho}{\partial r} . \tag{B3}$$

The hydrostatic equilibrium equation can then be written

$$-\frac{\partial}{\partial r}\left[\rho^{\Gamma-1}q\right] + \alpha \rho r = 0 . \tag{B4}$$

If we consider the limit of large radius $r$, we expect the density distribution to approach a pure power law so that $q \approx$ constant. In this case, the equilibrium equation reduces to

$$q^2(\Gamma - 1) + \alpha \rho^{2-\Gamma} r^2 = 0 . \tag{B5}$$

In order for this equation to have a solution, the second term must approach a constant value in the limit of large radius $r$ (recall that $\Gamma < 1$ here). This requirement implies that the power-law index $q$ must satisfy

$$q = \frac{2}{2 - \Gamma}, \qquad (B6)$$

a result which is valid for all equations of state with $\Gamma < 1$. Consequently, the index $q < 2$ and hence the integral determining the mass per unit length

$$\mu \equiv 2\pi \int_0^\infty \rho r \, dr, \qquad (B7)$$

is divergent. Thus, our initial claim is true: The mass per unit length diverges for any equation of state softer than isothermal.

## APPENDIX C: MASS PER UNIT AREA OF SHEETS

In this Appendix, we show that the mass per unit area for a molecular cloud sheet will be finite for any barotropic equation state stiffer than that of the logatropic limit. The simplest way to prove this claim is to introduce Lagrangian variables. In particular, we define the surface density $\sigma$ to be

$$\sigma(x) = \int_0^x \rho(x') \, dx'. \qquad (C1)$$

The equation of hydrostatic equilibrium can be written in the form

$$\frac{d\psi}{dx} + \frac{dP}{d\sigma} = 0, \qquad (C2)$$

where we now consider the surface density to be the dependent variable, consistent with our Lagrangian treatment. Next, we note that application of Gauss's law allows us to write the gravitational force in the form

$$\frac{d\psi}{dx} = \sigma. \qquad (C3)$$

Using this result in the hydrostatic equilibrium equation [C2] and integrating, we obtain the result

$$\frac{1}{2}(\sigma_\infty^2 - \sigma^2) = P(\sigma) - P_\infty, \qquad (C4)$$

where the subscript '$\infty$' indicates that the quantities are to be evaluated in the limit $x \to \infty$. The total mass per unit area of the sheet is simply $2\,\sigma_\infty$, where the factor of 2 arises because we must consider both sides of the sheet (i.e., both positive and negative values of $x$). As long as the equation of state is *stiffer* than the logatropic limit, then the pressure vanishes at spatial infinity (where the density $\rho \to 0$). In this case, we can evaluate equation [C4] at the midplane where $\sigma = 0$ to obtain the total mass per unit area $\sigma_{\text{tot}}$ of the molecular cloud sheet,

$$\sigma_{\text{tot}} = 2\sigma_\infty = 2[2P(\sigma = 0)]^{1/2} \, . \tag{C5}$$

Thus, our claim is true: The mass per unit area of any molecular cloud sheet will be finite for any barotropic equation of state stiffer than the logatropic limit.

## APPENDIX D: SEPARATION OF VARIABLES

We wish to separate the dependent variables in equation [4.5]. We assume a barotropic equation of state, $p = P(\rho)$, so that the pressure perturbation may be written in terms of the density perturbation.

$$p_1 = P'(\rho_0)\rho_1 \, . \tag{D1}$$

After substituting this into equation (2.10), we have

$$-\frac{\partial^2 \rho_1}{\partial t^2} + P_0' \nabla \rho_1 + \left[2P_0'' - \frac{1}{\rho_0}P_0'\right] \nabla \rho_0 \cdot \nabla \rho_1 + \nabla \rho_0 \cdot \nabla \psi_1 + A(\varpi)\rho_1 = 0 \, , \tag{D2}$$

where

$$A(\varpi) \equiv P_0'' \nabla^2 \rho_0 + P_0''' \left(\nabla \rho_0\right)^2 + 2\rho_0 \, , \tag{D3}$$

$P_0' \equiv P'(\rho_0)$, etc., and $\varpi$ represents the perpendicular coordinate. Let $\rho_1$ and $\psi_1$ have the forms

$$\rho_1(\varpi, z, t) = f(\varpi)g(z)h(t) \tag{D4}$$

$$\psi_1(\varpi, z, t) = \phi(\varpi)g(z)h(t). \tag{D5}$$

Equation [D2] then gives

$$\frac{h''(t)}{h(t)} = P_0' \left[\frac{g''(z)}{g(z)} + \frac{\nabla_\perp^2 f(\varpi)}{f(\varpi)}\right] + \left(2P_0'' - \frac{1}{\rho_0}P_0'\right)\frac{d\rho_0}{d\varpi}\frac{f'(\varpi)}{f(\varpi)} + \frac{d\rho_0}{d\varpi}\frac{\phi'(\varpi)}{f(\varpi)} + A(\varpi) = -\omega^2 \, , \tag{D6}$$

where the time dependence has been separated. (Recall that the equilibrium state is independent of both $t$ and $z$.) From this we conclude

$$h(t) = h_0 e^{\pm i\omega t}. \tag{D7}$$

From here we can procede to separate the $z$-dependence and the $\varpi$-dependence.

$$-\frac{g''(z)}{g(z)} = \frac{\nabla_\perp^2 f(\varpi)}{f(\varpi)} + \left(2\rho_0 \frac{P_0''}{P_0'} - 1\right)\frac{\rho_0'}{\rho_0}\frac{f'(\varpi)}{f(\varpi)} + \frac{\rho_0' \phi'(\varpi)}{P_0' f(\varpi)} + \frac{A(\varpi) + \omega^2}{P_0'} = k^2 \, , \tag{D8}$$

where $\rho_0' \equiv d\rho_0/d\varpi$. This gives

$$g(z) = g_0 e^{\pm ikz}. \tag{D9}$$

The remaining, non-trivial equation governs the structure of the perturbation perpendicular to the direction of propagation.

$$\nabla_\perp^2 f + \left(2\rho_0 \frac{P_0''}{P_0'} - 1\right) \frac{\rho_0'}{\rho_0} \frac{df}{d\varpi} + \frac{\rho_0' \phi'}{P_0'} + V(\varpi) = k^2 f, \tag{D10}$$

where

$$V(\varpi) = \frac{A(\varpi) + \omega^2}{P_0'}. \tag{D11}$$

## APPENDIX E: NUMERICAL METHODS

In this Appendix, we briefly describe the numerical techniques used in this paper. We apply the relaxation algorithm and routines of Press et al. (1992) to numerically solve equations [4.12] and [4.13]. By iterating a Runge-Kutta algorithm, we construct an approximate solution for large wavenumbers $k$ and use this solution as the initial "guess" for the relaxation algorithm. Then the relaxed solution for one value of $k$ is used as the initial guess for relaxation at a smaller value of $k$. In this fashion, we iteratively find solutions for smaller and smaller values of $k$, until adopted numerical parameters no longer provide an accurate solution. We use a separate numerical algorithm for the case $k = 0$.

As usual, this algorithm involves representing the solution functions on a mesh of points. Here we use a mesh of equally spaced points. Say the number of points is $m + 1$, and the spacing between points is $h$. Then $\varpi_j = jh$, for $j = 0, 1, 2, \ldots, m$. We apply boundary conditions at $\varpi_0 = 0$ and at $\varpi_m = mh \gg 1$. We wish the final grid point at $mh$ to be effectively at spatial infinity. So, when a solution is reached, the slope of the density at the "outer" boundary is checked. If it is larger than desired, indicating that the solution has not had sufficient room to decay, the quantity $h$ is increased, and the relaxation procedure is repeated.

# FIGURE CAPTIONS

Figure 1. The radial density profile of the static equilibria of hydrostatically supported filaments with various equations of state. The solid curve represents the isothermal filament; the short dashed curve, the purely logotropic filament; the dashed curve, a filament with a small turbulent pressure; the long dashed curve, a filament with a large amount of turbulence.

Figure 2. The vertical density profile of static equilibria of hydrostatically supported slabs with various equations of state. The solid curve represents the isothermal slab; the short-dashed curve, the purely logotropic slab; the dashed curve, a slab with a small turbulent pressure; the long-dashed curve, a slab with a large amount of turbulence.

Figure 3. The dispersion relations for a molecular cloud filament for various equations of state, $p = \rho + \kappa \log(\rho)$. The abscissa is the wave number multiplied by the effective sound speed at the center of the filament. The ordinate is the square of the frequency. The value of $\kappa$ is specified for each curve in the legend. Notice the effect upon the location of the zero and the minimum due to the variation in $\kappa$.

Figure 4. A sample of the radial density eigenfunctions $f_k(r)$. Figure 4a shows eigenfunctions for the isothermal filament. Eigenfunctions for a turbulent ($\kappa = 10$) filament are shown in Figure 4b.

Figure 5. Cross-sectional images of perturbed filaments. Density contours indicate $\rho = 0.0, 0.2, 0.4, \ldots, 1.2.$. Arrows indicate velocity. (a) A propagating wave in an isothermal filament; (b) A propagating wave in a turbulent ($\kappa = 10$) filament; (c) The fastest growing instability in an isothermal filament; (d) The fastest growing instability in a turbulent filament.

Figure 6. The ratio of the clump spacing $\lambda_{\text{fast}}$ to the half-width at half-max of the filament, $R_{\text{HWHM}}$. Points are shown for various values of turbulence parameter $\kappa$.

Figure 7. The growth rate $|\omega_{\text{fast}}|$ of the fastest growing mode of the filament. Points are shown for various values of the turbulence parameter $\kappa$.

Figure 8. A comparison of the numerical dispersion relation for the isothermal cylinder to that obtained from the charge density theory (Adams et al. 1994) using a Yukawa potential. The solid curve with points represents the dispersion relation obtained numerically in this work. The dashed curve shows a fit of the one-dimensional Yukawa dispersion relation to the numerical results.

Figure 9. The dispersion relations for the slab for various equations of state, $p = \rho + \kappa \log(\rho)$. The abscissa is the wave number multiplied by the effective sound speed at the center of the filament. The ordinate is the square of the frequency. The value of $\kappa$ is specified for each curve in the legend. Notice the effect upon the location of the zero and the minimum due to the variation in $\kappa$.

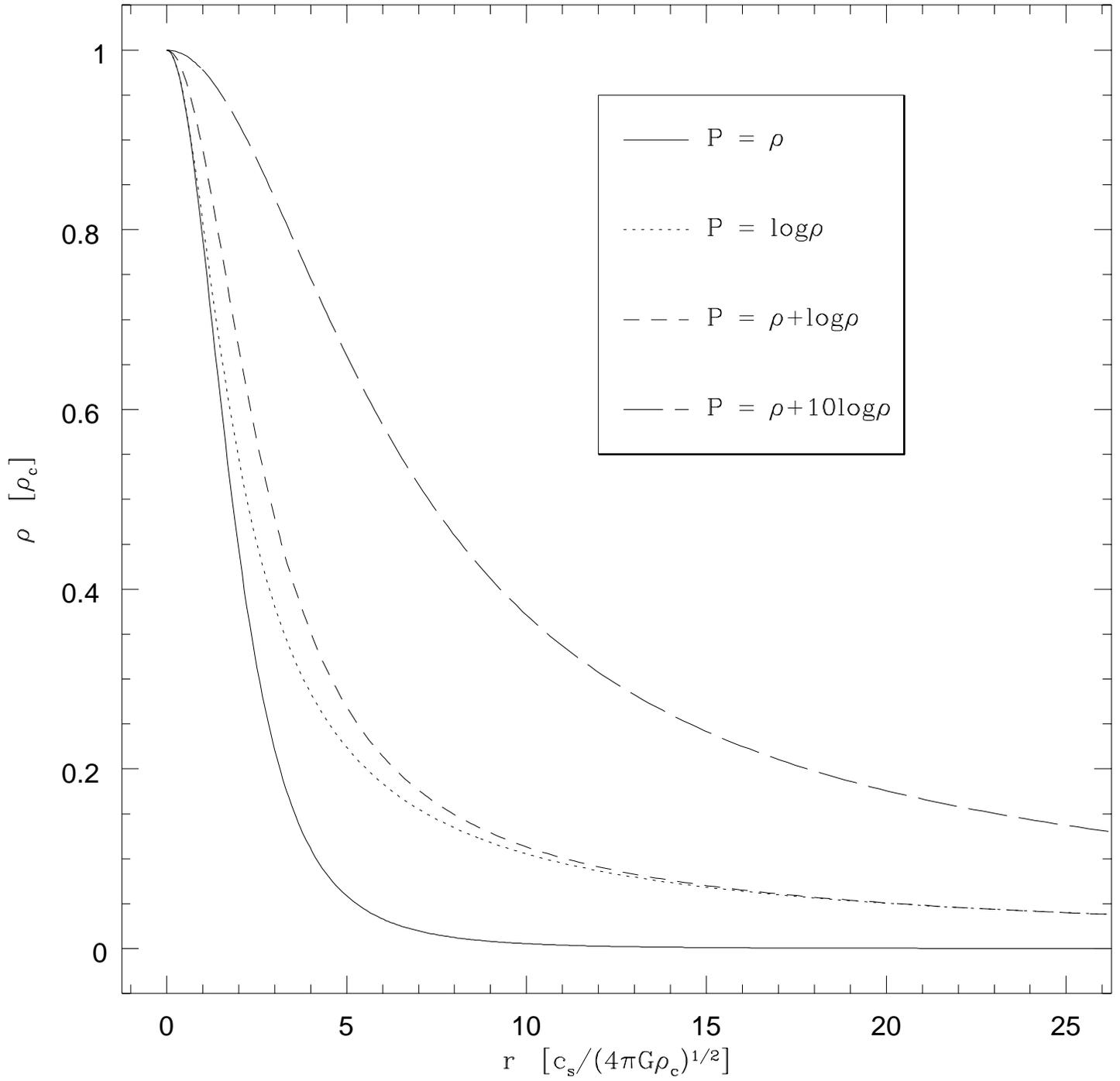

Figure 1

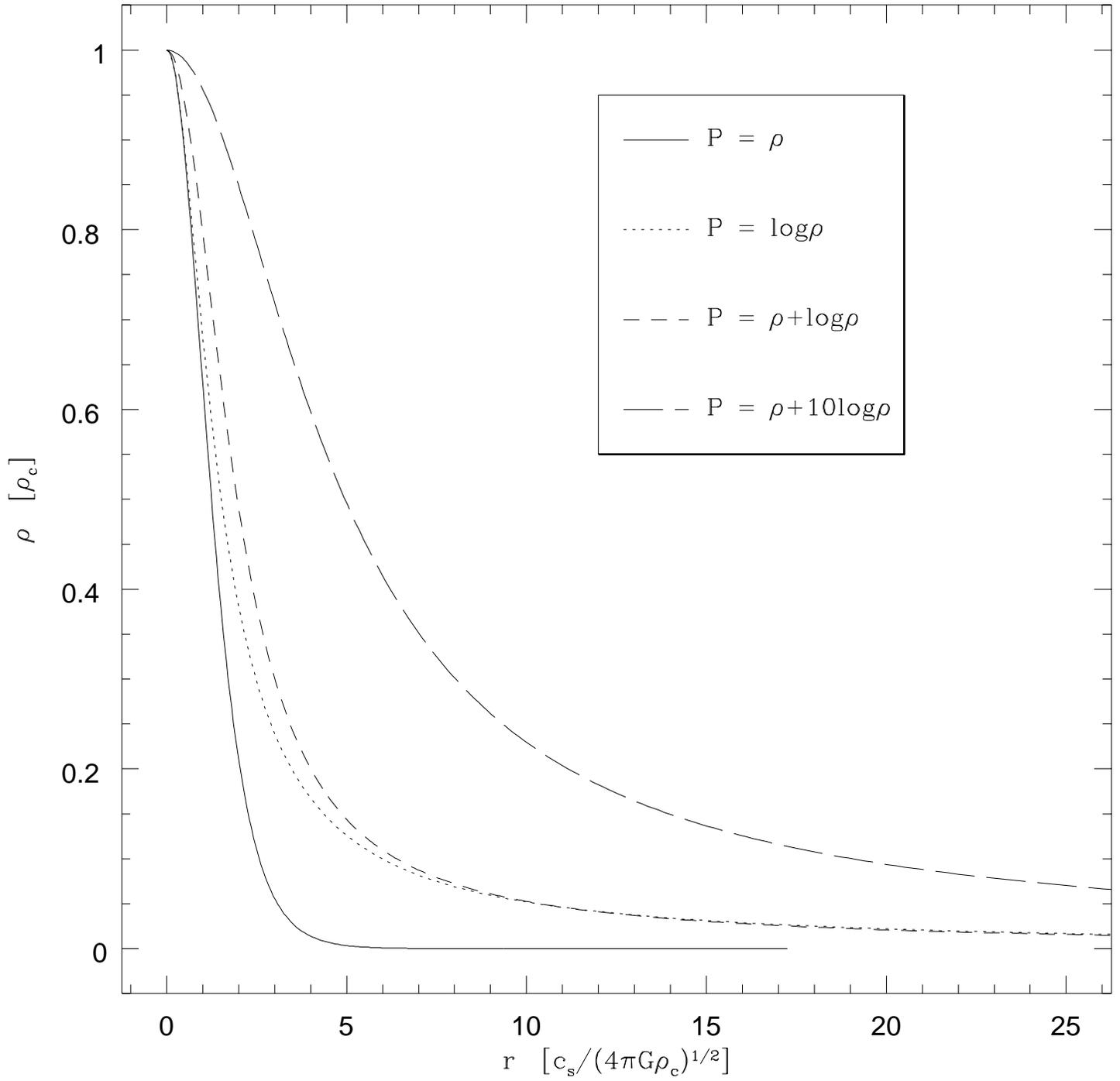

Figure 2

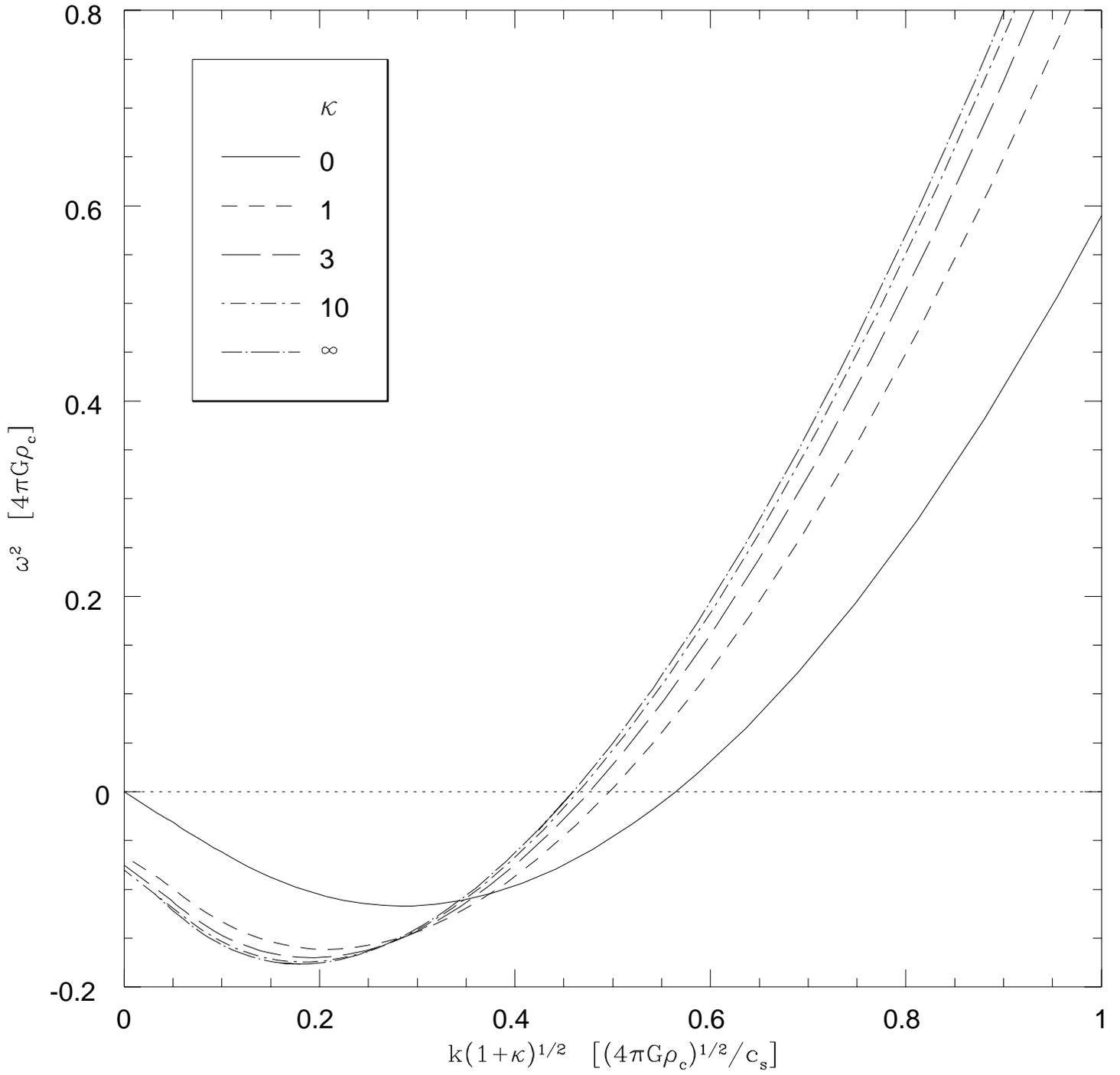
Figure 3

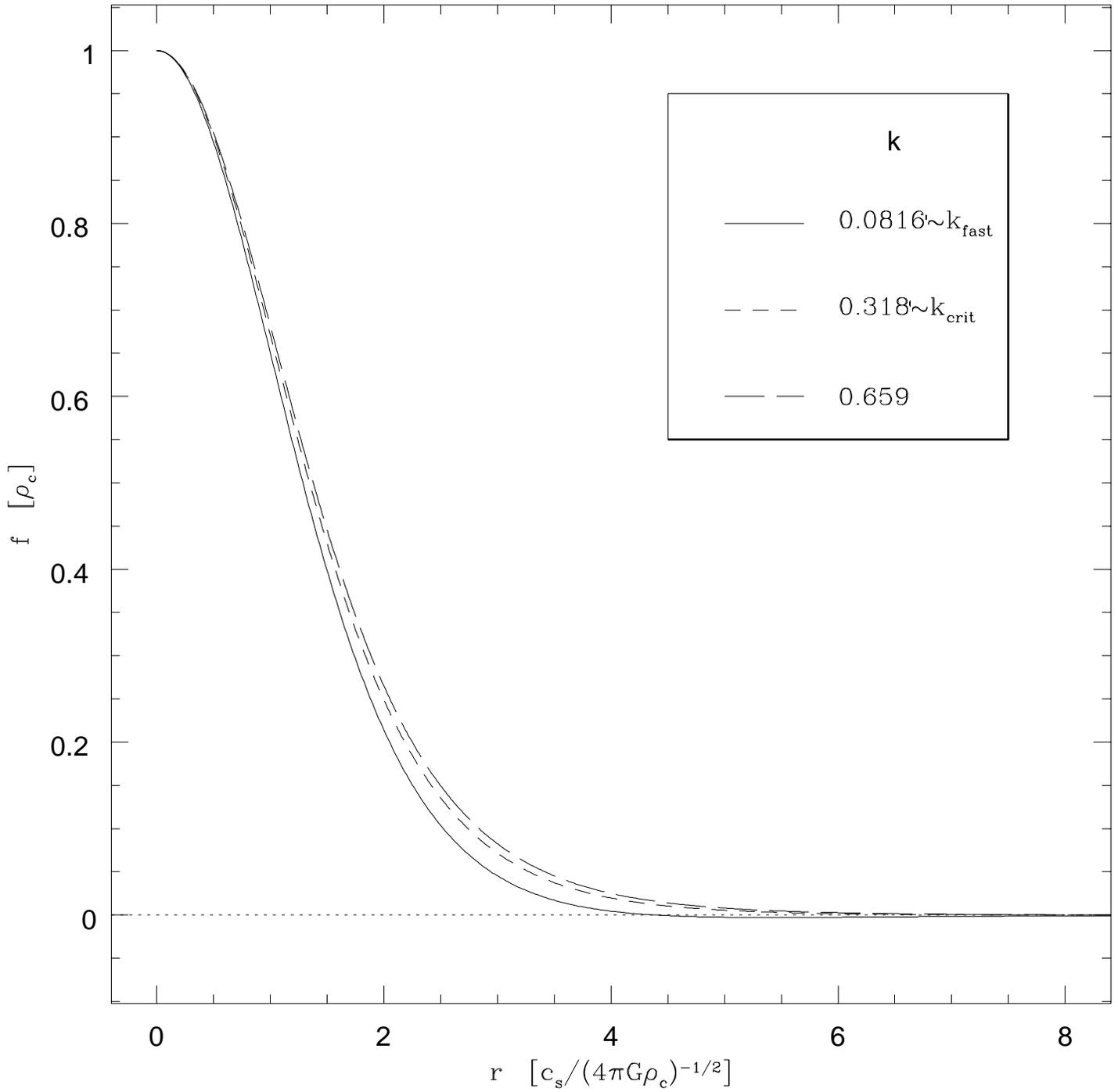
Figure 4a

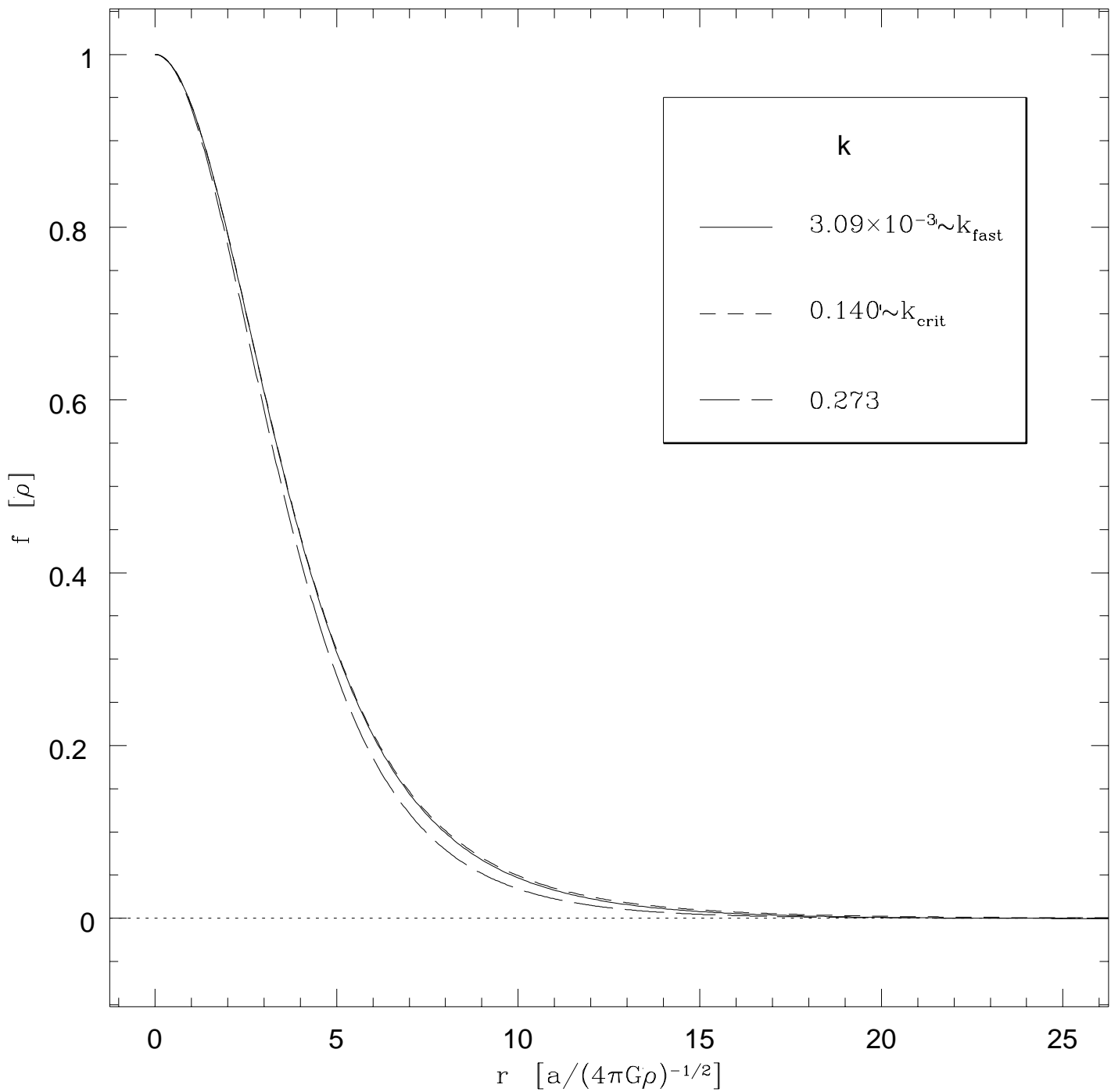

Figure 4b

Cross-sections of Perturbed Cylinders

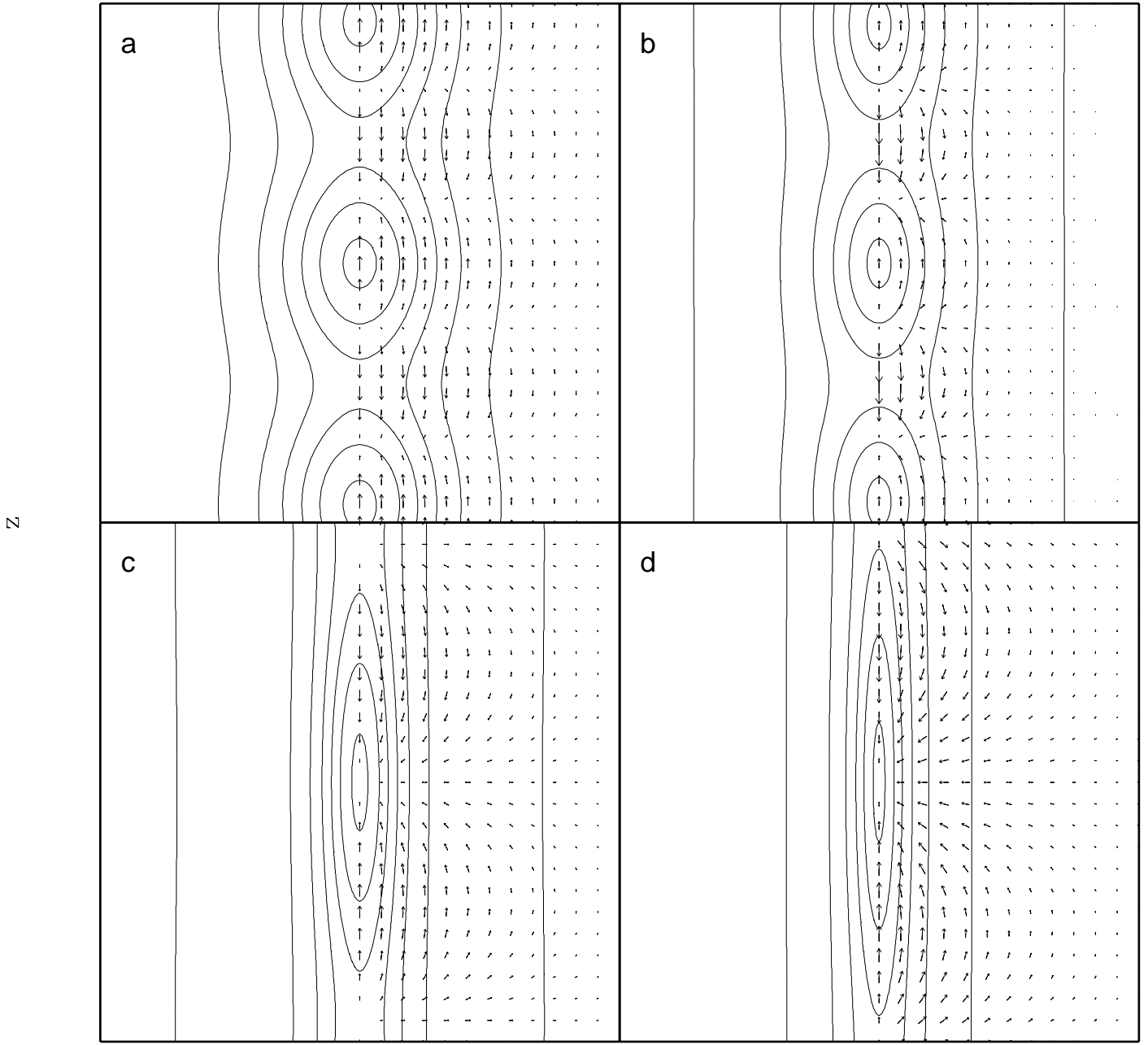

Figure 5

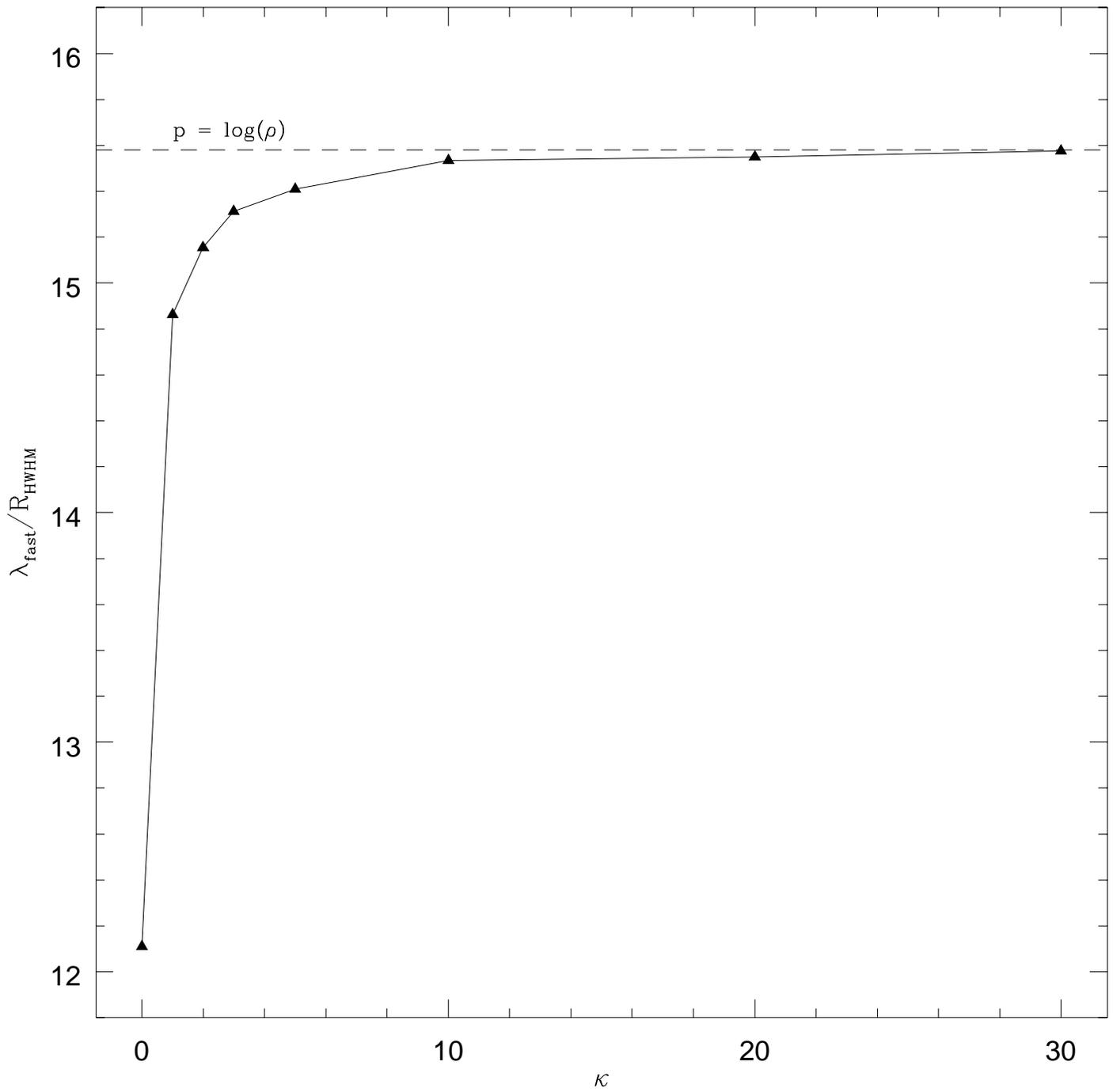

Figure 6

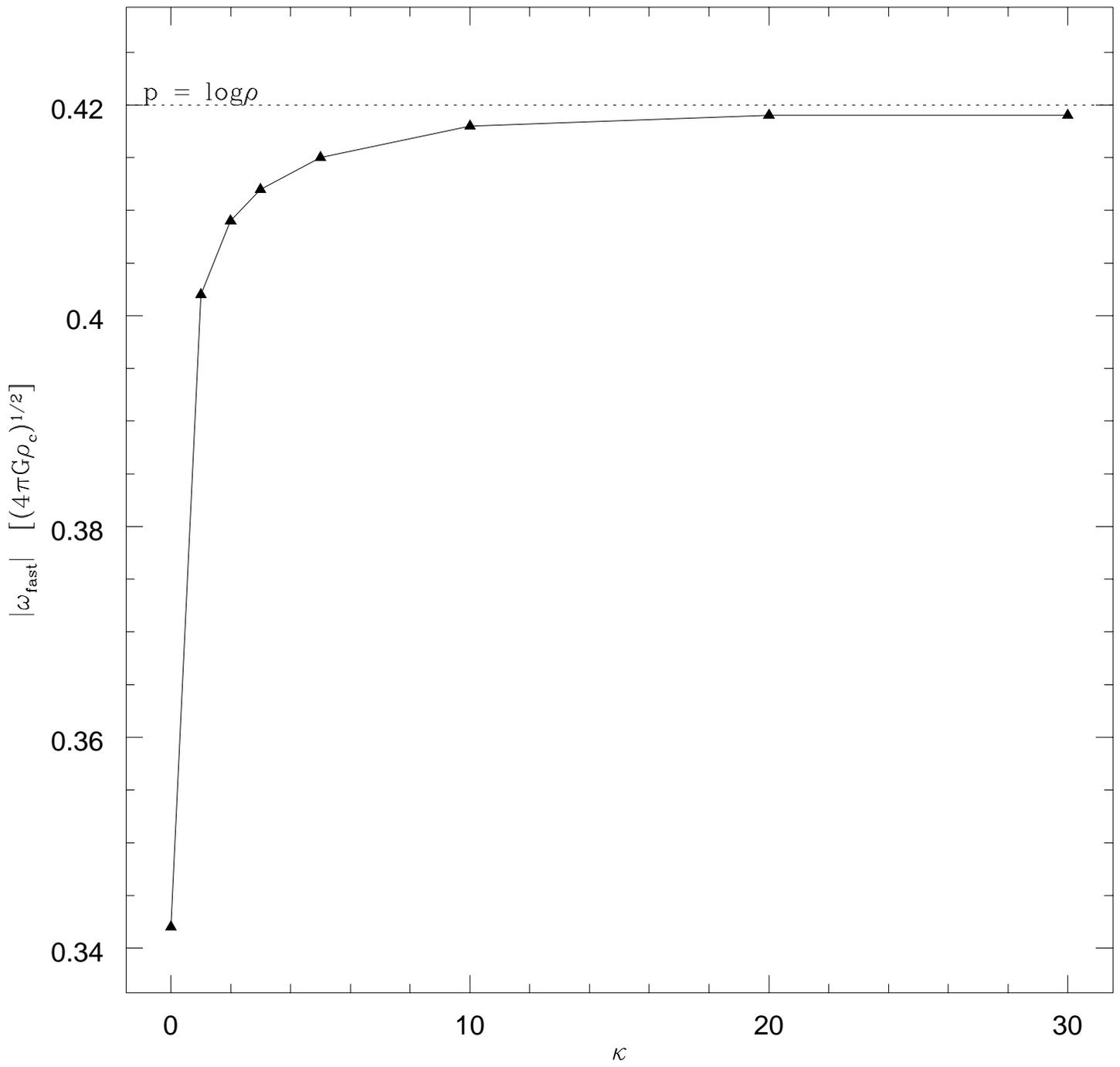

Figure 7

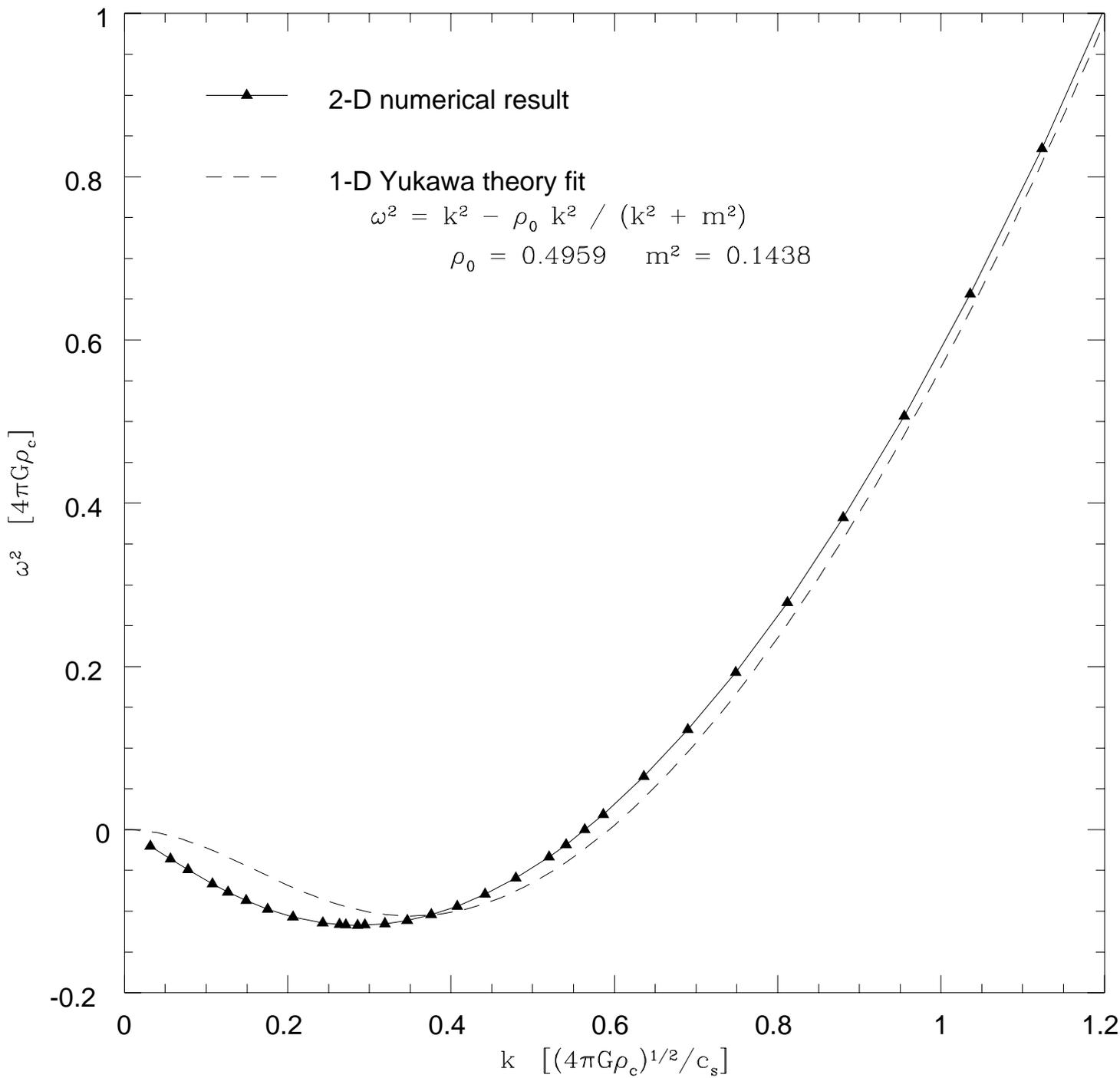

Figure 8

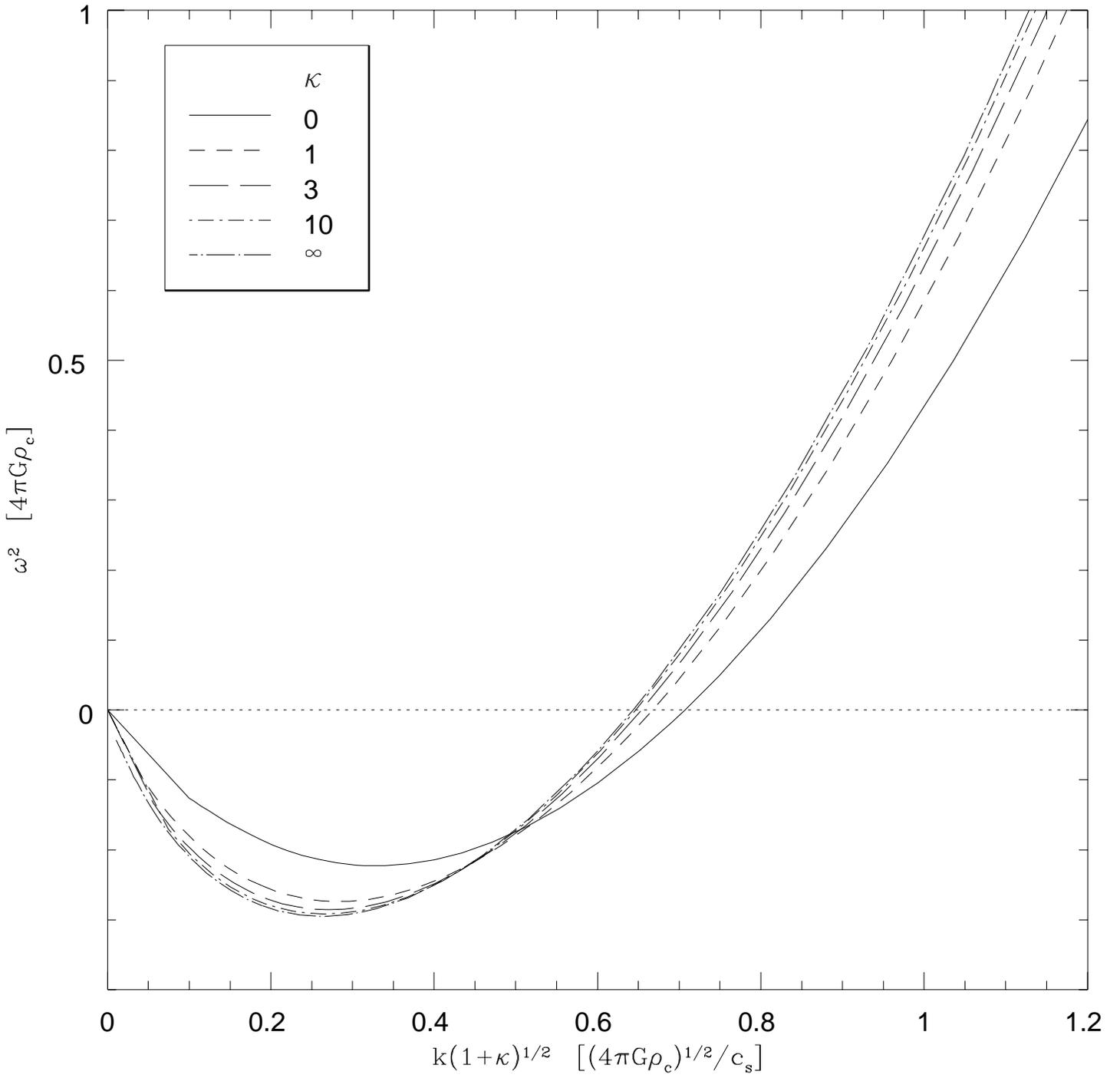

Figure 9